\documentstyle[preprint,aps]{revtex}

\newcommand{\be}{\begin{equation}}
\newcommand{\ee}{\end{equation}}
\newcommand{\bea}{\begin{eqnarray}}
\newcommand{\eea}{\end{eqnarray}}

\begin{document}

\begin{titlepage}
\begin{center}
\vskip .2in
\hfill
\vbox{
    \halign{#\hfil         \cr
           hep-th/9704189 \cr
           SU-ITP-97-14 \cr
           April 1997    \cr
           }  
      }   
\vskip 0.5cm
{\large \bf Supersymmetry Enhancement of D-p-branes and M-branes}\\
\vskip .2in
{\bf R. Kallosh}\footnote{e-mail address:kallosh@physics.stanford.edu} {\bf and
J. Kumar}
\footnote{e-mail address:jkumar@leland.stanford.edu}\\
\vskip .25in
{\em
Department of Physics,
Stanford University, Stanford, California 94305 USA \\}
\vskip 1cm
\end{center}
\begin{abstract}
We examine the supersymmetry  of classical D-brane
and
M-brane configurations and explain the  dependence of Killing spinors on
coordinates. We find that one
half supersymmetry is broken in the bulk and  that
supersymmetry near the D-brane horizon is restored for $p\leq 3$, for solutions
in the stringy frame, but only for $p=3$ in the10d canonical frame.  We study
the  enhancement for the case of four
intersecting
D-3-branes in  10 dimensions and the implication of this for the
size of the infinite throat of the near horizon geometry in non-compactified
theory. We found some indications of universality of near horizon geometries
of various
intersecting  brane configurations.
\vskip 0.5cm
 \end{abstract}
\end{titlepage}
\newpage

\section{Introduction}

The implications of the enhancement of supersymmetry of certain classical
solutions in
supergravity  near the horizon have been studied mostly in 4d and in 5d
\cite{Gibbons,KalPeet,GibTown,FKS,FK1,CFGK,arvind}.
In these cases, one finds
that some supersymmetries are broken in the bulk, but that the breaking
becomes weaker as one approaches the horizon.  In these dimensions
the extreme
black
holes with non-vanishing entropy, proportional to the area of the
horizon are
available.
The  enhancement of supersymmetry  near the horizon
of such black holes was instrumental for the study of the  entropy
of such
black holes \cite{FKS} and of a tension of a magnetic string
\cite{CKRRSW}. In
higher dimensions some $p$-branes and $M$-branes are
known to have an enhancemnt of supersymmetry near the horizon
\cite{GibTown}.
No such study has been performed for the $D$-branes \cite{Polch}.
One of the
purposes
of this paper is to fill in this gap.

The enhancement of supersymmetry near the horizon of intersecting and/or
overlapping branes
was also not studied before. In principle one may expect to find
some form of
supersymmteric attractors there as suggested in \cite{BalaLeigh}.
As the first step in
this direction we will find out what happens for the 4 intersecting
D-3-branes near the horizon. We will find out that near the horizon the
geometry of 3-3-3-3 solution is the same as of 6-2-2-2 and of 0-4-4-4
solutions.

The understanding of enhancement has been based largely on the
behavior of
the geometry and the dilaton near the horizon.
In all  known cases of {\it enhancement of supersymmetry}  the near
horizon
geometry is regular (in any frame) and of the form $adS_{p+2}
\times S^{d-p-2}$ and the dilaton is regular
\cite{Gibbons,KalPeet,GibTown,FKS,FK1,CFGK,arvind}.

In all  known cases of {\it absence of enhancement of
supersymmetry} near the
horizon the geometry is regular in stringy frame and of the form $M_{p+2}
\times S^{d-p-2}$ with the dilaton blowing up linearly near the
horizon in the
inertial frame of the Minkowski space $M_{p+2}$ \cite{KalPeet,GibTown}.

We will find out that for the D-p-branes that the situation is not as
simple since
in the stringy frame the near horizon geometry is not regular apart
from $p=3$.
It is actually conformal to a regular geometry of the type
$adS_{p+2}\times
S^{8-p}$ for all cases but $p=5$ (where the conformal geometry is
$M_{7}\times
S^{3}$). The conformal factor which brings $p\neq 3$ branes from the
stringy frame to the one with regular geometry is proportional to
$r^{p-3} $ and therefore is not a canonical 10d frame either. Also
the dilaton
is not regular at the horizon except for $p=3$. In view of this the
issue of
enhancement of supersymmetry near the horizon for the D-branes is
not on the
same footing as in all cases studied before. Therefore we will
simply proceed
with explicit evaluation of the supersymmetry transformations near
the horizon
in two natural frames: the stringy one and the canonical one.

Thus our main goal is to examine the supergravity transformations of various
classical solutions, to determine  whether or not they exhibit restoration of
supersymmetry at the horizon. It may be useful to mention here the following
general feature of Killing spinors. One can predict their dependence on
coordinates  for static solutions, based on supersymmetry algebra. Assume that
we have a geometry with some time component of the metric $g_{tt}(x)$ where $x$
are space coordinates. Killing spinors of unbroken supersymmetry
usually are found as a product of the function of space times the constant
spinor $\epsilon_0$.
\begin{equation}
\epsilon (x) =K(x)  \epsilon_0
\end{equation}
This can be understood as follows. The commutator of two supersymmetries has to
produce a translation. For static configurations the translation in time
direction has to be a  Killing vector. Introducing a vielbein $e_a{}^\mu$  we
have
\begin{equation}
(\bar \epsilon \Gamma^\mu \epsilon') { \partial  \over \partial  x^{\mu} } =
(\bar \epsilon \Gamma^0 \epsilon')  e_0{}^t \; { \partial  \over \partial  t
}=(\bar \epsilon_o  \Gamma^0 \epsilon_o ')  K^2 e_0{}^t \; { \partial  \over
\partial  t }=  { \partial  \over \partial  t }
\end{equation}
Thus
\begin{equation}
 K^2 e_0{}^t= K^2 g_{tt}^{-1/2} =1 \qquad \Longrightarrow \qquad K= (
g_{tt})^{1/4}
\end{equation}
and therefore the dependence of the Killing spinor has to be
\begin{equation}
\epsilon (x) =( g_{tt})^{1/4} \epsilon_0
\label{killing}\end{equation}
This indeed is the case for all static configurations.

In section 2, we will explicitly calculate the supersymmetry
transformations
of the classical fields of a D-p-brane background in the string metric of
N=2 D=10 supergravity.  We expect to find that, away from the
horizon, the
classical
solution is invariant under $1\over 2$ of all supersymmetry
transformations.
We will
then see if any of the supersymmmetries are restored near the
horizon.  In
section 3,
we will perform the same calculations for the 10d canonical frame
solutions for
a
D-p-brane background.  In section 4, we will investigate M-brane
classical
backgrounds of
11d supergravity and demonstrate the enhancement of supersymmetry near the
horizon of the 2-brane and the 5-brane.  In section 5, we will study the effect
of T-duality on
supersymmetry
in the bulk and near the horizon.  We will study the enhancement of
supersymmetry for solutions corresponding to intersecting D-branes in 10
dimensions in
section 6. In section 7 we will present some near horizon
geometries of the
different configurations of branes and discuss the universality issues.

\section{D-p-branes in the String Frame}

The supersymmetry transformations of the dilatino and gravitino
fields in the presence of a $p+2$ form gauge field strength in N=2 D=10
supergravity (IIA or IIB) are given by \cite{strframe}
\begin{equation}
\delta \psi_{\mu} = \partial_{\mu} \epsilon - {1\over 4}
\omega_{\mu} ^{ab}
\gamma_{ab}
\epsilon + {(-1)^p \over 8(p+2)!} e^{\phi} F_{\mu_1 ... \mu_{p+2}}
\gamma^{\mu_1 ...
\mu_{p+2}} \gamma_{\mu} \epsilon^{'} _{(p)}
\end{equation}
\begin{equation}
\delta \lambda = \gamma^{\mu} (\partial_{\mu} \phi) \epsilon + {3-p\over
4(p+2)!}
e^{\phi} F_{\mu_1 ... \mu_{p+2}} \gamma^{\mu_1 ... \mu_{p+2}}
\epsilon^{'}_{(p)}
\end{equation}
\begin{equation}
\epsilon_{(0,4.8)} ^{'} = \epsilon
\qquad
\epsilon_{(2,6)} ^{'} = \gamma_{11} \epsilon
\qquad
\epsilon_{(-1,3,7)} ^{'} = \imath \epsilon
\qquad
\epsilon_{(1,5)} ^{'} = \imath \epsilon^{\ast}
\end{equation}
where $\epsilon $ is a 32-component spinor, and $\omega$ is the spin
connection given by
\begin{equation}
\omega_{\mu} ^{ab} = -e^{\nu [a}(\partial_{\mu} e_{\nu} ^{b]} -
\partial_{\nu}
e_{\mu}
^{b]}) - e^{\rho [a} e^{\sigma b]} (\partial_{\sigma} e_{c \rho})
e_{\mu} ^c
\qquad
g_{\mu \nu} = e_{\mu} ^a e_{\nu} ^b \eta_{ab}
\end{equation}
The classical solution for the metric and fields of D-p-branes in the
string metric is given by
\begin{equation}
ds^2 = H^{-{1\over 2}} dx^2 _{(p+1)} - H^{{1\over 2}} dx^2 _{(9-p)}
\qquad
F_{01...pi} = \partial_{i} H^{-1}
\qquad
e^{2 \phi} = H^{-{p-3\over 2}}
\end{equation}
\begin{equation}
H = 1+\left({c\over r}\right)^{7-p}
\qquad
r^2 = x^2 _{(p+1)} +...+x^2 _{9}
\end{equation}
where the fermionic fields vanish (and consequently, so do the variations
in the graviton, dilaton, and gauge field strength).

For this solution, the spin-connection is then given by
\begin{equation}
\omega_r ^{\hat i \hat s} = {\partial_{i} H\over 4H^{{3\over 2}}}
\delta_{r \hat s}
\qquad
\omega_k ^{\hat i \hat j} = {\partial_{j} H\over 4H} \delta_{k \hat i}
\qquad
r,s \in \lbrace 0...p \rbrace
\qquad i,j \in \lbrace p+1 ... 9 \rbrace
\end{equation}
where $ \hat s $ is an index in the flat tangent space.

{}From this, we see that the supersymmetry transformations are now
given by
\begin{equation}
\delta \lambda = {(3-p)(\partial_i H)\gamma^i \over 4H^{5\over 4}}
[\epsilon +
\gamma_0 ... \gamma_p \epsilon^{'}]
\end{equation}
\begin{equation}
\delta \psi_r = \partial_r \epsilon + {(\partial_i H)\over 8H^{3\over 2}}
\gamma^i
\gamma_r [\epsilon + \gamma_0 ... \gamma_p \epsilon^{'}] = \hat
\nabla_{r}
\epsilon
\end{equation}
\begin{equation}
\delta \psi_i = \partial_i \epsilon -\sum_{i \neq j}
\left({(\partial_j H)\over
8H}
\gamma^i \gamma^j \right)
[\epsilon + \gamma_0 ... \gamma_p \epsilon^ {'}] -\left({\partial_i
H \over
8H}\right)
\gamma_0 ...
\gamma_p \epsilon ^ {'} = \hat \nabla_{i} \epsilon
\end{equation}
Clearly, these vanish for all $r$ if the following conditions are
satisfied
\cite{strframe}
\be
\label{spincon}
\epsilon + \gamma_0 ... \gamma_p \epsilon ^{'} = 0
\ee
\be
\label{field}
\epsilon = H^{-{1\over 8}} \epsilon_0 \ .
\ee
The solutions $\epsilon $ are Killing spinors.
The dependence on space coordinates here is found in agreement with the
prediction in (\ref{killing}), since here $H^{-{1\over 8}}= (g_{tt})^{1/4}$.
The spinor condition projects out half of the degrees of freedom of
the Killing
spinor for all D-p-brane backgrounds.  This amounts to saying that the
classical solutions to D-p-branes are preserved under half of the
supersymmetry
transformations.  Thus, D-p-branes break half of supersymmetry away
from the
brane.

The next task is to examine the transformations near the horizon.  We
know that for small $r$ and for $p<7$,
\begin{equation}
H \propto r^{p-7}
\end{equation}
Therefore, the transformations (viewed in the flat tangent space) will
vary with $r$ as
\begin{equation}
\delta \lambda , \delta \psi_r, \delta \psi_i \propto r^{3-p\over 4}
\end{equation}
Thus, for $ p<3 $, the classical solutions of the D-p-brane near
the horizon
will be invariant under all supersymmetry transformations near the
brane.  For
$7>p>3$, the dilatino
field is not invariant unless (\ref{spincon}) is satisfied.  Thus,
supersymmetry is half broken even at the horizon.  We find that for
$p=3$, the
dilatino field is invariant, while the gravitino field is not.  But the
gravitino field is not
gauge-invariant, so we must examine the transformation of the
gauge-invariant
gravitino
field strength in the flat tangent-space.  This transformation is
given by the
generalized
curvature tensor $(R_{ab})_{\alpha}{}^{ \beta}  $
\begin{equation}
\psi_{\mu \nu} =  \hat \nabla_{[\mu} \psi_{\nu ]}
\qquad
\delta \psi_{\mu \nu} = \hat \nabla_{[\mu} \delta \psi_{\nu]} = [\hat
\nabla_{\mu} ,
\hat \nabla_{\nu}] \epsilon
\end{equation}
\begin{equation}
(\delta \psi_{ab})_{\alpha} = e^{\mu} _a e^{\nu} _b [\hat
\nabla_{\mu} , \hat
\nabla_{\nu}] \epsilon
= [\hat \nabla_a , \hat \nabla_b ] \epsilon = (R_{ab})_{\alpha}{}^{
\beta}
\epsilon_{\beta}
\end{equation}
Note that the supercovariant derivative, $\hat \nabla $ is the sum of the
covariant
derivative and a term involving the gauge field strength.  Thus, the
generalized
curvature tensor is the sum of the Riemann curvature tensor and
terms involving
the gauge field strength.
By plugging in the supersymmetry transformations given above, we find
the following integrability condition
\begin{equation}
[\hat \nabla_{\hat r} , \hat \nabla_{\hat s}] \epsilon = 0
\end{equation}
\begin{equation}
[\hat \nabla_{\hat r} , \hat \nabla_{\hat i}] \epsilon =\lbrace
-\partial_i
\left({\partial_k H\over 8H^{3\over 2}}\right)
\gamma^k \gamma_r +{(\partial_k H)^{2}\over 32H^{5/2}} \gamma_i \gamma_r
\rbrace (\epsilon + \gamma_0 ... \gamma_p \epsilon_{(p)} ^{'})
\end{equation}
\begin{eqnarray}
[\hat \nabla_{\hat i} , \hat \nabla_{\hat j} ]\epsilon = {1\over
8}\lbrace
-H^{-{1\over 2}}\left(\partial_i \left({
\partial_l H\over H}\right) \gamma^j \gamma^l  - \partial_j
\left({\partial_k
H\over H}\right) \gamma^i
\gamma^k \right) + {(\partial_k H)^{2}\over 4H^{5\over 2}}
[\gamma^i \gamma^j ]
\nonumber\\
-{(\partial_i H)(\partial_l H)\over 2H^{5\over 2}} \gamma^l \gamma^j +
{(\partial_j H)(\partial_k H)\over 2H^{5\over 2}} \gamma^k \gamma^i
\rbrace
(\epsilon
+ \gamma_0 ... \gamma_p \epsilon_{(p)} ^{'})
\end{eqnarray}
Again, this vanishes for all $r$ if (\ref{spincon}) is satisfied.  We
also find that each term in the generalized curvature tensor
is proportional to $ r^{3-p\over 2}$.
We can see this more easily by noting that
\begin{equation}
\hat \nabla_r , \hat \nabla_i \propto r^{3-p\over 4}   , r \rightarrow 0
\end{equation}
Thus, the generalized curvature tensor vanishes, term by term, as $r
\rightarrow 0$
for $p<3$.  In particular,
the Riemann tensor in the tangent space vanishes as $r \rightarrow 0$ for
$p<3$.  Since
the tangent space is flat, this implies that the curvature scalar
(which is
found by contracting the Riemann tensor with the metric) also
vanishes.  Note
that this supersymmetry enhancement occurs for all of the cases where the
dilaton
blows up as $r \rightarrow 0$, since
\be
e^{2\phi} \sim r^{(7-p)(p-3)\over 2}
\qquad
\phi \sim {(7-p)(p-3)\over 4 } \ln{ r}
\ee
For $p=3$ in the limit as $r \rightarrow 0$, we find that the
curvature does
not vanish.
However, the various terms in the field strength transformations
cancel each
other.  Thus, the generalized curvature vanishes, and full
supersymmetry is
restored in the
 $p=3$ case as well.  This
matches our geometric understanding of the situation, as shown in
\cite{GibTown}.
As $r \rightarrow 0$, the D-3-brane metric tends
to $adS_5 \times S_5 $.  This calculation was performed in the
string frame,
and thus
enhancement only occurs when $r\ll l_s $.  It is not clear how one
interprets
this from a string-theoretic point of view.

\section{Supersymmetry of D-p-branes in the 10d Canonical Metric}

In the 10d canonical metric, the supersymmetry transformations of
the dilatino
and gravitino fields are given by \cite{DuffLu}
\be
\delta \lambda = {1\over 2\sqrt 2 }(\nabla_M \phi ) \gamma^M \gamma^{11}
\epsilon
+{(3-p)\over 8\sqrt 2 (p+2)! } e^{{(3-p)\over 4}\phi} \gamma^{M_1
... M_{p+2}}
F_{M_1 ... M_{p+2}} \epsilon_{(p)} ^{'}
\ee
\be
\delta \psi_M = \nabla_M \epsilon + {(3-p)3^{p-1} \over 32(p+2)! }
e^{{3-p\over
4}
\phi}\left(\gamma_M ^{M_1 ... M_{p+2}} - {(7-p)(p+2)\over
p+1}\delta_M ^{M_1}
\gamma^{M_2 ... M_{p+2}}\right)F_{M_1 ... M_{p+2}} \epsilon_{(p)} ^{'}
\ee
\be
\epsilon_{(0)} ^{'} = \imath \epsilon
\qquad
\epsilon_{(1,2)} ^{'} = \epsilon
\ee
and the classical solution for D-p-branes is given by
\be
ds^2 = H^{p-7\over 8}(dt^2 - dx_1 ^2 - ... -dx_p ^2 ) - H^{p+1\over
8}(dx_{p+1}
^2 + ... +dx_9 ^2 )
\qquad
H = 1+\left({c\over r}\right)^{7-p}
\ee
\be
F_{01...pi} = -{\partial_i H\over H^2 }
\qquad
e^{2\phi} = H^{3-p\over 2}
\ee
We will find that, in this case, we can determine all of the
information we need about
enhancement from the dilatino variation, which reduces to
\be
\delta \lambda = ({3-p\over 8\sqrt 2}){(\partial_i H)\over H}
H^{-{p+1\over
16}}
\gamma^i \gamma^{11} [\epsilon +\gamma^{11} \gamma_0 ... \gamma_p
\epsilon_{(p)} ^{'}]
\ee
As expected, we find that, away from the horizon, the solution is
preserved
only when $\epsilon $ obeys the spinor condition
\be
\epsilon + \gamma^{11} \gamma_0 ... \gamma_p \epsilon_{(p)} ^{'} = 0
\ee
As $r \rightarrow 0 $, we find that
\be
\delta \lambda \longrightarrow r^{-{(3-p)^2 \over 16}}
\ee
Thus, we find that , if $p\neq 3$, supersymmetry is not enhanced in the 10d
canonical frame.    For $p=3$, the dilaton is regular and the 10d canonical
frame is the
same as the string frame, wherein we have already determined that supersymmetry
is enhanced at the horizon.

It seems at first strange that supersymmetry should
appear in some cases to be enhanced in the string metric, while not in the 10d
canonical metric.  But this should not be too suprising, since each metric
measures
the supersymmetry breaking on a different scale.  One notes that the
integrability condition is very closely related to the curvature, which is
clearly different depending on whether one uses the string metric or canonical
metric.  It may be that the extent to which some supersymmetries are broken
goes to zero
when measured on the string scale, but not when measured on the 10d Planck
scale.  One
must then ask which scale is appropriate for asking questions regarding
supersymmetry
enhancement.  This may be a question whose answer depends on M-theoretic
considerations.

In any case, these ambiguities do not affect us when dealing with D-3-branes,
for which the
string frame and canonical frame are identical.  We will see in section 7 that
we can find
a class of solutions which exhibit near-horizon $adS$ geometry.  These
solutions are all
equivalent to a configuration involving only intersecting 3-branes, for which
there is no
ambiguity.

\section{Supersymmetry of Classical Solutions of M-branes}
Using the 11d canonical metric, we find the supersymmetry
transformation
\cite{Cremmer,DuffLPS}
\begin{equation}
\label{11trans}
\delta \psi_{\mu} = \partial_{\mu} \epsilon - {1\over 4}
\omega_{\mu} ^{ab}
\gamma_a
\gamma_b \epsilon + {i\over 288}\left(\gamma_{[ \mu} \gamma^{\alpha}
\gamma^{\beta}
\gamma^{\gamma} \gamma^{\delta ]} - 8\delta_{\mu} ^{\alpha}
\gamma^{[ \beta}
\gamma^{\gamma} \gamma^{\delta ]}\right) F_{\alpha \beta \gamma \delta}
\epsilon
\end{equation}
The supermembrane classical field configuration is given by
\begin{equation}
ds^2 = H^{-{2\over 3}}(dt^2 - dx_1 ^2 - dx_2 ^2 ) - H^{1\over
3}(dx_3 ^2 + ...
+
dx_{10} ^2)
\qquad
H = 1 + \left({c\over r}\right)^{8-p}
\end{equation}
\begin{equation}
F_{012i} = -{\partial_i H \over H^2}
\end{equation}
We thus find the following transformations for the gravitino field
strength
\begin{equation}
[\hat \nabla_{\hat r} , \hat \nabla_{\hat s} ] = 0
\end{equation}
\begin{equation}
[\hat \nabla_{\hat r} , \hat \nabla_{\hat i} ] = -[H^{1\over 6}
\partial_i
({\partial_k H\over
6H^{3\over 2}}) \gamma^k - {(\partial_k H)^2\over 36H^{7\over 3}}
\gamma_i
+{(\partial_k H)(\partial_i H)\over 36H^{7\over 3}} \gamma^k ]
\gamma_r (1 + i
\gamma^0 \gamma^1 \gamma^2)
\end{equation}
\begin{eqnarray}
[\hat \nabla_{\hat i} , \hat \nabla_{\hat j} ] = -[H^{-{1\over
3}}(\partial_i
({\partial_l H\over
12H})\gamma^l \gamma_j - \partial_j ({\partial_k H\over
12H})\gamma^k \gamma_i
)
+{(\partial_k H)^2\over 72H^{7\over 3}} [\gamma^i , \gamma^j ] +
{(\partial_k
H)
(\partial_j H)\over 36H^{7\over 3}} \gamma_i \gamma^k \nonumber \\ -
{(\partial_i H)(\partial_l H)\over 36H^{7\over 3}} \gamma_j \gamma^l](1 +
i\gamma^0 \gamma^1
\gamma^2)
\end{eqnarray}
In the bulk, these variations vanish only when the spinor condition is
satisfied.  As
$ r \rightarrow 0$ we see that the variations vanish for any
$\epsilon $.  Thus
supersymmetry
is half-broken in the bulk, but is enhanced at the horizon of the
M-2-brane in
the 11d canonical
frame.

The classical solution for the M-5-brane is given by
\be
ds^2 = H^{-{1\over 3}} (dt^2 + dx_1 ^2 +...+ dx_5 ^2 ) - H^{{2\over
3}} (dx_6
^2
+ ... +dx_{10} ^2 )
\qquad
H=1+\left({c\over r}\right)^3
\ee
\be
F^{\alpha \beta \gamma \delta} = -\epsilon^{012345 \alpha \beta
\gamma \delta
\epsilon }
{\partial_{\epsilon} H\over H^2 }
\ee
Using Eq. (\ref{11trans}), and plugging in as before, we find
\be
[\hat \nabla_{\hat r} , \hat \nabla_{\hat s}] = 0
\ee
\be
[\hat \nabla_{\hat r}, \hat \nabla_{\hat i}] = [-H^{1\over 6} \partial_i
({\partial_k H\over
12H^{3\over 2}}) \gamma^k + {(\partial_k H)^2 \over 36H^{8\over 3}}
\gamma^i +
{(\partial_k H)(\partial_i H)\over 72H^{8\over 3}} \gamma^k ]\gamma_r (1+
i\gamma^6 ...
\gamma^{10} )
\ee
\begin{eqnarray}
[\hat \nabla_{\hat i} , \hat \nabla_{\hat j} ] = H^{-{2\over
3}}[\partial_i
\left({\partial_l H\over 6H}\right)\gamma^j \gamma^l
- \partial_j \left({\partial_k H\over 6H}\right) \gamma^i \gamma^k
+{(\partial_k H)^2
\over 18H^2 }[\gamma^i , \gamma^j ] \nonumber\\ + {(\partial_j
H)(\partial_k
H)\over
9H^2 }\gamma^i \gamma^k -{(\partial_l H)
(\partial_i H)\over 9H^2 }\gamma^j \gamma^l ](1 + i\gamma^6 ...
\gamma^{10} )
\end{eqnarray}
Again, this yields half supersymmetry breaking in the bulk, but as $r
\rightarrow 0$ the
variations vanish for all $\epsilon $.  Thus, we see supersymmetry
enhancement
for both
the 2-brane and 5-brane of M-theory in the 11d canonical frame. On
the basis of
the properties of the geometry $adS_4\times S^7$ for the 2-brane and
$adS_7\times S^4$  for the 5-brane near the horizon, the enhancement of
supersymmetry was studied in \cite{GibTown}.
Here we have in addition checked that the generalized curvature
vanishes near the horizon
and therefore there are no constraints on Killing spinors near the horizon.

\section{T-Duality}

It may at first seem odd that supersymmetry is enhanced for $p=3$
but not for
$p=4$.  One might expect that configurations
T-dual to the D-3-brane will exhibit the same supersymmetry at the
horizon.  In
particular, one
might expect the D-4-brane solution to also exhibit enhancement.
But one must
first note that the
4-brane and 3-brane solutions given above are not T-dual.  The
T-dual of the 4-brane solution \cite{bho} is given by
\be
ds^2 = H^{-{1\over 2}}(dt^2 - dx_1 ^2 -...-dx_3 ^2 ) - H^{{1\over 2}} (
dx_4 ^2 +...+ dx_9 ^2 )
\qquad
H = 1 + \left({c\over r}\right)^3
\ee
which depends on the harmonic function of the 4-brane.  But in any case,
\cite{bko}
showed that T-duality does not necessarily respect supersymmetry,
even in the
bulk.  They found
that the supersymmetry of a configuration is preserved only if one
dualizes
along a direction
on which the Killing spinor does not depend.  From Eq.
(\ref{field}) we see
that for a D-brane
solution,
the graviton field is only preserved if the Killing spinor depends on the
transverse coordinates.
When we dualize the 3-brane solution, we find that we break
supersymmetries at
the horizon
only if we increase the D-brane dimension, which occurs precisely when we
dualize
along directions upon which the Killing spinor depends.

\section{Near horizon supersymmetry of intersecting D-p-branes}

We consider the classical solution for four D-3-branes, pairwise
intersecting on one-branes.  For simplicity, we make the particular
choice of orientations (1 2 3),(3 4 5),(5 6 1) and (2 4 6).  We then
have the classical solution \cite{TseytSource,2255,Bala,gaunt}

\be
ds^2 = H^{-2}dt^2 - H^2 (dx_7 ^2 + dx_8 ^2 + dx_9 ^2 ) -(dx_1 ^2+ ...
+ dx_6 ^2)
\qquad
H = 1+{c\over r}
\ee
\be
\hat F_{0123i}=\hat F_{0345i}=\hat F_{0561i}=\hat F_{0246i} =
-{\partial_{i} H\over H^2 }
\qquad
F_{\alpha \beta \gamma \delta} = {1\over 2}(\hat F_{\alpha \beta
\gamma \delta}
+
\ast \hat F_{\alpha \beta \gamma \delta})
\ee
We have chosen the four 3-brane charges to be the same for simplicity,
although the argument holds for arbitrary charges.  When compactified
down to 4 dimensions, this solution is known to form a black hole with
$1\over 8 $ supersymmetry in the bulk, but $1\over 4 $ supersymmetry
near the horizon.  Using the previous methods, one derives the
supersymmetry transformations
\be
\delta \lambda = 0
\ee
\be
[\hat \nabla_{\hat 0} , \hat \nabla_{\hat i}] = [\partial_i
({\partial_k H\over
2H^3 }) \gamma_k + {(\partial_k H)^2 \over 2H^4 } \gamma_i
]\gamma_0 \Delta_1
\ee
\begin{eqnarray}
[\hat \nabla_{\hat i} , \hat \nabla_{\hat j} ] = [\partial_i ({\partial_l
H\over
2H}){\gamma_l \gamma_j \over H} -\partial_j ({\partial_k H\over
2H}){\gamma_k
\gamma_i \over H} +{1\over 2}{(\partial_k H)^2 \over H^4 }[\gamma_i
, \gamma_j
]
-{(\partial_l H)(\partial_i H)\over H^4 }\gamma_l \gamma_j \nonumber\\
+{(\partial_k H)
(\partial_j H)\over H^4 }\gamma_k \gamma_i ]\Delta_1 - {3\over
8}[{(\partial_k
H)^2
\over 2H^4 }[\gamma_i , \gamma_j ] - {(\partial_l H )(\partial_i
H)\over H^4 }
\gamma_l \gamma_j + {(\partial_k H )(\partial_j H )\over H^4
}\gamma_k \gamma_i
]
\Delta_2
\end{eqnarray}
\be
[\hat \nabla_{\hat r} , \hat \nabla_{\hat s} ] = 0
\ee
\be
[\hat \nabla_{\hat 0} , \hat \nabla_{\hat r} ] = -{(\partial_k H
)^2 \over
16H^4 } \gamma_0
\gamma_r [\pm (1+\gamma_1 \gamma_2 \gamma_4 \gamma_5) \pm
(1+\gamma_3 \gamma_4
\gamma_6 \gamma_1) \pm (1+\gamma_5 \gamma_6 \gamma_2 \gamma_3 )]
\ee
\begin{eqnarray}
[\hat \nabla_{\hat r} , \hat \nabla_{\hat i} ] = -[\partial_i
({\partial_k
H\over 2H^2 })
{\gamma^k \over H} - {(\partial_k H )^2 \over 2H^4 }\gamma_i
+{(\partial_i H
)(\partial_l H )\over
2H^4 } \gamma_l ]\Gamma \gamma_r (\pm {i\Gamma \over 4}\gamma_0 \gamma_1
\gamma_2
\gamma_3 \pm
{i\Gamma \over 4}\gamma_0 \gamma_3 \gamma_4 \gamma_5 \nonumber\\
\pm {i\Gamma
\over 4}
\gamma_0 \gamma_5 \gamma_6 \gamma_1 \pm {i\Gamma \over 4}\gamma_0
\gamma_2
\gamma_4 \gamma_6 ) + [{(\partial_k H)^2 \over 16H^4 }\gamma_i \gamma_r -
{(\partial_i H)(\partial_l H)\over 8H^4 } \gamma_l \gamma_r ] \Delta_2
\end{eqnarray}
\be
\Delta_1 = 1 + {i\Gamma \over 4}\gamma_0 \gamma_1 \gamma_2 \gamma_3
+{i\Gamma
\over 4}\gamma_0 \gamma_3 \gamma_4 \gamma_5 +{i\Gamma \over 4} \gamma_0
\gamma_5
\gamma_6 \gamma_1 +{i\Gamma \over 4} \gamma_0 \gamma_2 \gamma_4 \gamma_6
\qquad
\Gamma = {1+\Gamma_{11} \over 2}
\ee
\be
\Delta_2 = 1 + {1\over 3} \gamma_1 \gamma_2 \gamma_4 \gamma_5 +
{1\over 3}
\gamma_3
\gamma_4 \gamma_6 \gamma_1 +{1\over 3} \gamma_5 \gamma_6 \gamma_2
\gamma_3
\ee

These  transformations involve two different types  of spinor
projector combinations.  The first is a sum of 4 projectors
\be
4\Delta_1 = (1+i\Gamma_{\alpha}) + (1+i\Gamma_{\beta}) +
(1+i\Gamma_{\gamma})
+(1+i\Gamma_{\delta})
\ee
\be
\Gamma_{\alpha} = \gamma_0 \gamma_a \gamma_b \gamma_c
\ee
whose gamma matrix indices are in the directions along the 3-branes.
Only three of these projectors are independent, however, since
\be
\epsilon = -i\Gamma_{\alpha} \epsilon
= -i\Gamma_{\beta} \epsilon
= -i\Gamma_{\beta} \epsilon \longrightarrow \epsilon = -i
(-\Gamma_{\alpha}
\Gamma_{\beta} \Gamma_{\gamma}) \epsilon =-i\Gamma_{\delta} \epsilon
\ee
Thus, this combination of spinor projectors breaks supersymmetry to
$1\over 8$.  The second projector combination is a sum of three
projectors of the form
\be
\pm (1+ \Gamma_{12}) \pm  (1+\Gamma_{13}) \pm (1+\Gamma_{23})
\qquad
\Gamma_{\alpha \beta} = \Gamma_{\alpha} \Gamma_{\beta}
\ee
where the term with gamma matrices is a product of any two of the
analogous
terms in the first projector combination.  By noting that
\be
\Gamma_{\alpha} \Gamma_{\beta} = \Gamma_{\beta} \Gamma_{\alpha}
\qquad
\Gamma_{\alpha} \Gamma_{\beta} \Gamma_{\gamma} = -\Gamma_{\delta}
\ee
we see that there are only three possible ways to form a term of a
projector of
this
type.  We can in fact write this projector in the more symmetric form
\be
1+{1\over 6} \sum_{i<j} \Gamma_{ij}
\ee
We find that of the three projectors, only two are independent.
Therefore,
this projector combination preserves $1\over 4$ of supersymmetry.  We see
\be
\epsilon = -i\Gamma_{\alpha} \epsilon , \epsilon = -i\Gamma_{\beta}
\rightarrow \epsilon = -\Gamma_{\alpha \beta}
\ee
Thus, if $\epsilon $ survives a projector combination of the first type,
it also satisfies the second.  Therefore, the $1\over 8$ supersymmetry
preserved by the first projector is a subset of the $1\over 4$
supersymmetry preserved by the second.

The transformations in the bulk preserve $1\over 8 $ of the total
supersymmetry.  As
$r \rightarrow 0 $,
\be
[\hat \nabla_{\hat 0} , \hat \nabla_{\hat i} ] \longrightarrow 0
\ee
\be
[\hat \nabla_{\hat i} , \hat \nabla_{\hat j} ] \longrightarrow
-{3\over 8c^2
}({[\gamma_i ,
\gamma_j ]\over 2 } - {x^l x^i \over r^2 }\gamma_l \gamma_j +{x^j
x^k \over r^2
} \gamma_k
\gamma_i ) \Delta_2
\ee
\be
[\hat \nabla_{\hat r} , \hat \nabla_{\hat i} ] \longrightarrow
({\gamma_i \over
16c^2 } -
{x^i x^l \over 8r^2 c^2 } \gamma_l ) \gamma_r \Delta_2
\ee
which preserves $1\over 4 $ supersymmetry, as expected.  Thus, we
are able
to recover the supersymmetry of a 4-dimensional black hole, both in the
bulk and at the horizon, using a 10-d supergravity calculation of
intersecting D-brane solutions.

\section{Universality of the near horizon geometries}

We may consider the 3-3-3-3 solution of intersecting D-branes with  harmonic
functions
$H_{\alpha}=1 + {r_{\alpha} \over  r}$ with 4 different parameters
$r_{\alpha}$ for each harmonic function.  We then find the following geometry
\cite{Bala,2255,gaunt}

\begin{eqnarray}
ds^2 = (H_{\alpha} H_{\beta} H_{\gamma} H_{\delta} )^{-{1\over 2}} dt^2 -
(H_{\alpha} H_{\beta} H_{\gamma} H_{\delta} )^{-{1\over 2}} (dx_7 ^2 + ...
+dx_8 ^2 ) - \nonumber\\ (H_{\alpha} H_{\beta} H_{\gamma} H_{\delta} )^{1\over
2}\left({dx_1 ^2 \over H_{\alpha} H_{\beta} } +...+ {dx_6 ^2 \over H_{\gamma}
H_{\delta} }\right)
\end{eqnarray}

The meaning the $r_{\alpha}$ is that it measures the size of the
throat and
the volume of the sphere of $adS_{5}\times S^{5}$, which is the
near horizon
geometry of a single 3-brane.

For the intersecting solution of four  3-branes the near horizon
canonical
geometry   has a nice form: the geometry is that of an anti-de
Sitter space
times a circle and times an Euclidean space: $adS_{2}\times S^{2} \times
E^6$.
\begin{eqnarray}
ds^2 &= &{r^2 \over \sqrt {r_{\alpha} r_{\beta}  r_{\gamma}
r_{\delta}  }}
dt^2 - { \sqrt {r_{\alpha} r_{\beta}  r_{\gamma}  r_{\delta}  }
\over r^2}
dr^2 -  \sqrt {r_{\alpha} r_{\beta}  r_{\gamma}  r_{\delta}  }  \; d^2
\Omega_2
\nonumber\\
&-& \sqrt {r_{\alpha} r_{\beta}  r_{\gamma}  r_{\delta}  }  \left(
{dx_1 ^2
\over r_{\alpha} r_{\beta}} + ...
+ {dx_6 ^2 \over r_{\gamma}  r_{\delta}}\right)
\end {eqnarray}
The size of the infinite throat is now  given by the inverse scalar
curvature
of
the $adS_{2}$ geometry \cite{CFGK}
\begin{equation}
2\pi \sqrt {r_{\alpha} r_{\beta}  r_{\gamma}  r_{\delta}  } = {4\pi
\over R}
\end{equation}
The flat 6-dimensional Euclidean geometry is not of the standard
form $ds^2 =
\delta_{ij} d \tilde x d \tilde x$, as each direction has to be
rescaled with
different constant parameters to bring the coordinates $x^i$ to
$\tilde x^i$.
This expression upon compactification will become the entropy of
the black
hole. From the point of view of the 10d geometry it is the size of
the near
horizon throat of $adS_{2}$.

However, we also find evidence that there is in fact a class of
solutions which exhibit the same near-horizon geometry.  For example,
consider the 0-4-4-4 solution, with the three 4-branes pairwise intersecting
on 2-branes.  The metric is given by \cite{BalaLL}

\begin{eqnarray}
ds^2 = (H_0 H_{4 \alpha} H_{4 \beta} H_{4 \gamma} )^{-{1\over 2}} dt^2 -
(H_0 H_{4 \alpha} H_{4 \beta} H_{4 \gamma} )^{-{1\over 2}} (dx_7 ^2 + ...
+dx_8 ^2 ) - \nonumber\\ (H_0 H_{\alpha} H_{\beta} H_{\gamma} H_{\delta}
)^{1\over 2}\left({dx_1 ^2 \over H_{4\alpha} H_{4\beta} } +...+ {dx_6 ^2 \over
H_{4 \beta} H_{4 \gamma} }\right)
\end{eqnarray}
with  $H_0 = 1 + {r_0 \over r} $ the harmonic function associated with the
0-brane, and $H_{4 \alpha} = 1 + {r_{4 \alpha} \over r} $ associated with
the three 4-branes.  Near the horizon, this geometry approaches

\begin{eqnarray}
ds^2 = {r^2 \over \sqrt{r_0 r_{4 \alpha } r_{4 \beta } r_{4 \gamma }}}dt^2
-{\sqrt{r_0 r_{4\alpha} r_{4\beta} r_{4\gamma}}\over r^2 }dr^2
-\sqrt{r_0 r_{4\alpha} r_{4\beta} r_{4\gamma}} d\Omega_2 \nonumber\\
-\sqrt{r_0
r_{4\alpha} r_{4\beta} r_{4\gamma}}\left({dx_1 ^2 \over r_{4\alpha} r_{4\beta}}
+...+{dx_6 ^2 \over r_{\beta} r_{\gamma}}\right)
\end{eqnarray}
As before, the size of the $adS_2$ throat is given by
\be
2\pi \sqrt{r_0 r_{4\alpha} r_{4\beta} r_{4\gamma}} = {4\pi \over R}
\ee

Another example is the 6-2-2-2 solution, with the 2-branes intersecting
at a point, and all 2-branes embedded within the 6-brane.  Its geometry
is given by \cite{BalaLL}

\begin{eqnarray}
ds^2 = (H_6 H_{2\alpha} H_{2\beta} H_{2\gamma})^{-{1\over 2}} - (H_6
H_{2\alpha} H_{2\beta} H_{2\gamma})^{1\over 2}(dx_7 ^2 +...+dx_9 ^2 )
\nonumber\\ -
({H_{2\alpha} H_{2\beta} H_{2\gamma}\over H_6 })^{1\over 2}\left({dx_1 ^2
\over H_{2\alpha}} +...+ {dx_6 ^2 \over H_{2\gamma}}\right)
\end{eqnarray}
where $H_6 = 1+{r_6 \over r}$ is the harmonic function for the 6-brane
and $H_{2\alpha} = 1+{r_{2\alpha} \over r}$ are associated with the three
2-branes.  Near the horizon, this geometry tends to

\begin{eqnarray}
ds^2 = {r^2 \over \sqrt{r_6 r_{2\alpha} r_{2\beta} r_{2\gamma}}} dt^2 -
{\sqrt{r_6 r_{2\alpha} r_{2\beta} r_{2\gamma}}\over r^2}dr^2 - \sqrt{r_6
r_{2\alpha} r_{2\beta} r_{2\gamma}} d\Omega_2 \nonumber\\ - \sqrt{
r_{2\alpha} r_{2\beta} r_{2\gamma} \over r_6 }\left({dx_1 ^2 \over r_{2\alpha}}
+...+ {dx_6 ^2 \over r_{2\gamma}}\right)
\end{eqnarray}
This is an $adS_2 \times S^2 \times E^6$ geometry with $adS_2$ throat volume
given by
\be
2\pi \sqrt{r_6 r_{2\alpha} r_{2\beta} r_{2\gamma}} = {4\pi \over R}
\ee

It is possible that this indicates a universality of near horizon black hole
geometries in 10 and 11 dimensions.  Specifically, it may be that all classical
solutions for intersecting branes in 10 and 11 dimensions which preserve
$1\over
8$ supersymmetry in the bulk and which have constant (but non-zero) curvature
at
the horizon (in the tangent space) exhibit a near horizon geometry of the form
$adS_l \times S^m \times E^{n}$.  Imagine that we have a metric of the form

\be
ds^2 = H^y dx_{(t)} ^2 + H^{x_1} dx_1 ^2 +...+ H^{x_n} dx_n ^2
\qquad
H \rightarrow \left({q\over r}\right)^{-z} , r \rightarrow 0
\ee
where $z=d-2$, $d$ is the number of overall transverse directions, and the
$x_t$ are the coordinates in those directions.
By examining the spin-connection and covariant derivatives in a straightforward
manner, one can show that

\be
\hat \nabla_{\mu} \propto r^{{yz\over 2}-1}
\ee
Thus, curvature is only constant when $zy = 2$.  This (and the demand
of $1\over 8$ supersymmetry) form powerful constraints on the solutions which
can be examined.  Suppose we only consider conventional solutions, by which we
mean solutions for intersecting branes such that each term of the metric is the
product of the appropriate terms in the metric solutions of the individual
branes.
One-eighth supersymmetry requires the presence of at least
three branes.  In ten dimensions, we may assume that $e^{\phi} =1$, since that
seems typical when the curvature at the horizon is constant but non-zero.  In
that case, the string frame is the same as the canonical frame, and we find
that
for a (conventional) solution with three branes, $y={3\over 2}$.  This does not
allow for an
integer value of $z$, and is thus unacceptable.  For a solution with four
branes,
we find we must have $y=2, z=1$, and we have 3 overall transverse directions.
In order to have $1\over 8$ supersymmetry from 4 branes, we must demand that
any
direction have either 0, 2, or 4 branes extend along it.  Quite clearly, this
will
give us a near horizon geometry of $adS_{2+x} \times S^2 \times E^{6-x}$.

For solutions in 11 dimensions
involving the intersection of three branes, we find that the
solutions given in
\cite{gaunt} for the $2\perp 2\perp 5$ and $2\perp 5\perp 5$ are
non-regular at
the
horizon.  The solution for $2\perp 2\perp 2$ yields a near horizon
geometry
(after
a simple coordinate transformation) of

\be
ds^2 = \left[ {2\delta  \over (r_{\alpha} r_{\beta}
r_{\gamma})^{1\over 3} }
\right]^2 dt^2 -\left[ {(r_{\alpha} r_{\beta} r_{\gamma})^{1\over
3} \over
2\delta}\right]^2 dr^2 -(r_{\alpha} r_{\beta} r_{\gamma})^{2\over
3} d\Omega_3
-(r_{\alpha} r_{\beta} r_{\gamma})^{1\over 3}
\left({dx_1 ^2 \over r_{\alpha}} + ... +{dx_6 ^2 \over r_{\gamma}}\right)
\ee

This is $adS_2 \times S^3 \times E^6$.  The solution for $5\perp
5\perp 5$ can
be transformed to a metric of the form

\begin{eqnarray}
ds^2 = \left[{r\over 2(r_{\alpha} r_{\beta} r_{\gamma})^{1\over
3}}\right]^2
(dt^2
-dx_{10} ^2 )
-\left[{2(r_{\alpha} r_{\beta} r_{\gamma})^{1\over 3}\over
r}\right]^2 dr^2 +
(r_{\alpha} r_{\beta} r_{\gamma})^{2\over 3} d\Omega_2 \nonumber\\
 -(r_{\alpha} r_{\beta} r_{\gamma})^{1\over 3} \left({dx_1 \over r_{\alpha}
r_{\beta}} +...+ {dx_6 \over r_{\beta} r_{\gamma}}\right)
\end{eqnarray}
This is $adS_3 \times S^2 \times E^6$, which, in 5 dimensions, is
the dual of
$adS_2 \times S^3 \times E^6$.

We also examine the intersections of the two 2-branes and two 5-branes in
11d. There are 2 parameters in the harmonic functions for the
2-branes $r_1,
r_2$ and two parameters in the harmonic functions for the
5-branes $\hat r_1, \hat r_2$.  For the 2-branes the parameters
measure the
size of the $adS_{4}$ throat of the near horizon geometry of a
single 2-brane.
For the 5-branes the parameters measure the size of the $adS_{7}$
throat of the
near horizon geometry of a single 5-brane. These parameters appear
in the near
horizon intersecting solution on an unequal footing, as opposed to
the cases
examined above.

The near horizon geometry of $2\perp 2 \perp 5 \perp 5$ \cite{TseytSource,2255}
in the canonical 11d
frame is  a product space of the type an anti-deSitter space times
a circle and
times an Euclidean space: $adS_{2}\times S^{2} \times
E^6 \times E^1$. Universality appears if one performs a
conformal
rescaling of the 11d metric with a constant parameter
\begin{eqnarray}
  \left({\hat r_1 \hat r_2 \over  r_1 r_2}\right) ^{1/6}ds_{11} ^2
&= & \left[
{r^2 \over \sqrt {r_{1} r_{2}  \hat r_{1}  \hat r_{}  }} dt^2 - { \sqrt
{r_{\alpha} r_{\beta}  r_{\gamma}  r_{\delta}  } \over r^2}   dr^2
-  \sqrt {r_{\alpha} r_{\beta}  r_{\gamma}  r_{\delta}  }  \; d^2
\Omega_2
\right] \nonumber\\
&-& \sqrt {r_{\alpha} r_{\beta}  r_{\gamma}  r_{\delta}  }  \left[
{dx_1 ^2
\over r_{\alpha} r_{\beta}} + ...
+ {dx_6 ^2 \over r_{\gamma}  r_{\delta}}\right]  +   \left({\hat
r_1 \hat r_2
\over  r_1 r_2}\right) ^{1/2} (dx_7)^2
\end {eqnarray}

In this form again we may recognize the size of the anti-deSitter
throat which
will measure the entropy of the black hole in 4d upon compactification.

Thus some indication of universality  comes out from this analysis
of geometries even before compactification.

\section{Conclusion}

We have found that, in the string frame, classical D-brane supergravity
solutions
preserve $1\over 2$ supersymmetry in the bulk, but preserve full
supersymmetry
at the horizon for $p\leq 3$.  In the 10d canonical frame, however,
supersymmetry
is enhanced only for $p=3$.  It seems that the next step is to
understand this
supersymmetry enhancement from the string theory point of view.

We have found supersymmetry enhancement for M-theory 2-branes and
5-branes, as
expected.
We have also found enhancement for the configuration of 4
D-3-branes, pairwise
intersecting on 1-branes.  This configuration, when compactified to 4
dimensions,
is known to give a black hole which exhibits the same enhancement.
We described the near horizon geometry of different configurations
in 10d and
11d and found some signatures of universal behaviour near the horizon
before compactification. In particular, it appears that all
solutions in 10
and 11 dimensions which preserve $1\over 8$
supersymmetry in the bulk and which have a regular (but non-zero) Riemann
tensor at the horizon exhibit a near-horizon geometry of the form
$adS_l \times
S^m \times E^n $.  The size of the anti-de Sitter throat then gives
the entropy
of the corresponding configuration, when compactified.

\acknowledgements

We would like to thank G. Horowitz, J. Rahmfeld,  A. Rajaraman  and Wing Kai
Wong
for essential
discussions.   The work of R. K. and of J. K. has been supported  by NSF
grant PHY-9219345. The work of J. K. has been also supported by the Department
of
Defense,
NDSEG Fellowship Program.

\end{document}